\documentclass[twocolumn,tighten]{aastex631}

\graphicspath{{./}{figures/}}
\usepackage{amsmath}

\usepackage{array,booktabs,makecell}
\usepackage{rotating}

\begin{document}

\title{Potential for life to exist and be detected on Earth-like planets orbiting white dwarfs}

\correspondingauthor{Caldon~T.~Whyte}
\email{cwhyte2021@my.fit.edu}

\author[0009-0006-9091-1706]{Caldon~T.~Whyte}
\affiliation{Department of Aerospace, Physics and Space Sciences, Florida Institute of Technology, Melbourne, FL 32901, USA}

\author[0000-0002-9390-955X]{L.~H.~Quiroga-Nu\~nez}
\affiliation{Department of Aerospace, Physics and Space Sciences, Florida Institute of Technology, Melbourne, FL 32901, USA}

\author[0000-0002-2685-9417]{Manasvi Lingam}
\affiliation{Department of Aerospace, Physics and Space Sciences, Florida Institute of Technology, Melbourne, FL 32901, USA}

\author[0000-0001-8764-1780]{Paola Pinilla}
\affiliation{Mullard Space Science Laboratory, University College London, Holmbury St Mary, Dorking, London, UK}

\begin{abstract}

With recent observations confirming exoplanets orbiting white dwarfs, there is growing interest in exploring and quantifying the habitability of temperate rocky planets around white dwarfs. In this work, the limits of the habitable zone of an Earth-like planet around a white dwarf are computed based on the incident stellar flux, and these limits are utilized to assess the duration of habitability at a given orbital distance. For a typical $0.6 M_\odot$ white dwarf an Earth-like planet at $\sim 0.012$ AU could remain in the temporally evolving habitable zone, maintaining conditions to support life, for nearly 7 Gyr. In addition, additional constraints on habitability are studied for the first time by imposing the requirement of receiving sufficient photon fluxes for UV-mediated prebiotic chemistry and photosynthesis. We demonstrate that these thresholds are comfortably exceeded by planets in the habitable zone. The prospects for detecting atmospheric biosignatures are also evaluated, and shown to require integration times on the order of one hour or less for ongoing space observations with JWST. 

\end{abstract}



\section{Introduction} \label{sec:intro}

The search for habitable planets has recently become very promising with advances in instrumentation provided by the James Webb Space Telescope (JWST), which has already begun to probe into the atmospheres of exoplanets, searching for signs of life \citep{Kempton2024, 2024NatAs...8..810T, Hammond2024, zi2024}. Future observations performed by the Nancy Grace Roman Telescope will look to further expand the list of exciting prospects and justify the development of even more powerful instruments that could give definite evidence for extrasolar life, such as the Habitable Worlds Observatory (HWO) \citep{Carrion2021, Tamburo2023, Harada2024}. There are also notable ground-based telescopes such as the Vera C. Rubin Observatory \citep{Krabbendam2024}, ESO's Extremely Large Telescope (ELT) \citep{Tamai2024}, and the Giant Magellan Telescope (GMT) \citep{Burgett2024} which are nearing completion and will be able to probe exoplanets at wavelengths ranging from near ultraviolet to mid-infrared. This ability to observe thousands of exoplanets provides the opportunity to prioritize systems with the highest chance of detecting life in the search for habitable exoplanets. 

Traditionally, Sun-like stars have been perceived as ideal hosts for habitable exoplanets, as the Sun is the only known star to host a habitable planet. It may end up being the case that this type of star is the best for supporting life \citep{Haqq2018, Lingam_Loeb2018, Lingam_Loeb_2021}, but the characteristics of other types warrant their consideration. This includes the long lifetimes and prevalence of red dwarfs \citep{Tarter2007, Shields2016} and brown dwarfs \citep{barnes2013habitable, Lingam_Loeb2019,lingam2020prospects}. A stellar type that seems to combine a fairly long lifetime with an effective temperature similar to Sun-like stars is a white dwarf. Approximately 97\% of all stars in the Milky Way will become a white dwarf, as it is the fate of any star with a mass below $\sim10 M_\odot$ \citep{Fontaine01, Althaus2010, Saumon2022}. Without nuclear fusion as the primary source of energy, they emit residual thermal energy and slowly cool. This cooling is closely connected to the internal structure of the star, and crystallization can slow this cooling for a few billion years through the release of latent heat \citep{Hansen2004, barnes2013habitable}. 

\begin{figure*}[htbp] 
    \centering 
    \includegraphics[width=\textwidth]{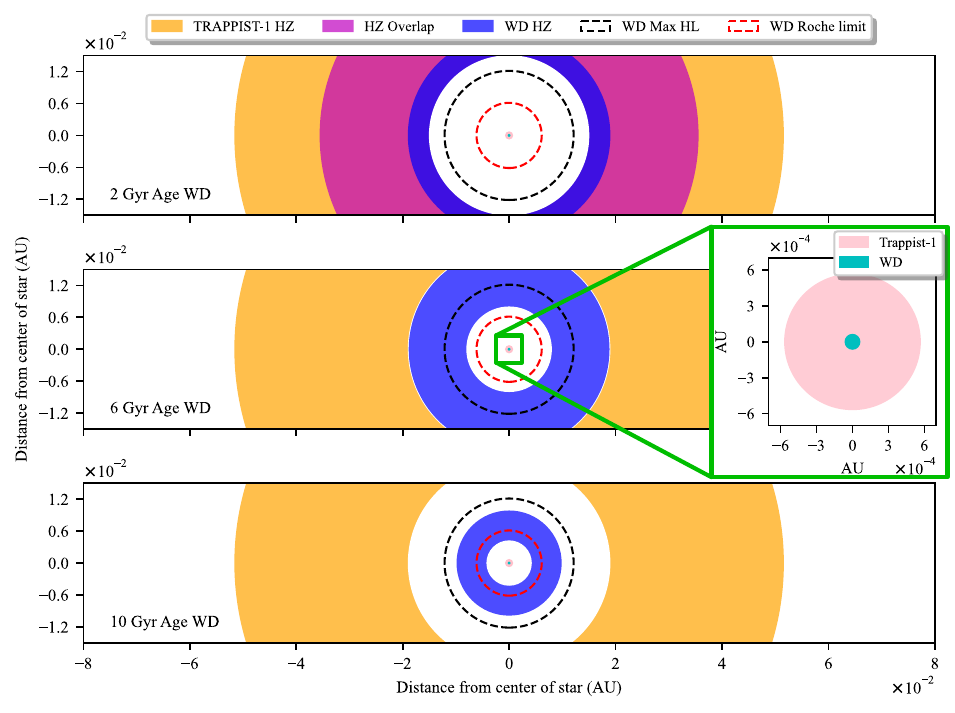}
    \caption{Habitable zone for a white dwarf (blue region) at ages of 2, 6, and 10 Gyr compared to the habitable zone of the red dwarf TRAPPIST-1 (orange region). The inset shows the size comparison of a typical $0.6 M_\odot$ white dwarf (cyan region) to that of TRAPPIST-1 (pink region). The optimistic habitable zone for TRAPPIST-1 is calculated from the recent Venus and early Mars limits from \citet{Kopparapu_2013} using the luminosity and effective temperature from \citet{VanGrootel2018}. The magenta region specifies where overlap of the two habitable zones occurs. The dashed black line shows the orbital radius at which an Earth-like planet around a $0.6 M_\odot$ white dwarf stays in the habitable zone for the maximum amount of time (i.e., maximum habitable lifetime), and the dashed red line shows the Roche limit (i.e., tidal disruption limit) for such a system.} 
    \label{fig:HZ_comp} 
\end{figure*} 

The process that leads to a star becoming a white dwarf is potentially detrimental to the fate of its planets. During post-main-sequence evolution, variations in the star's radius and mass create a chaotic system that would likely destabilize its planetary system \citep{kippenhahn2012stellar, Lamers2017, pinsonneault2023stellar}. The detection of pollution in up to 50\% of white dwarfs shows the unlikelihood of survival of planets as it implies that planet accretion must be common \citep{Zuckerman10, Koester14, Farihi_2016}. However, there are various mechanisms that possibly explain how planets can survive this transition period \citep{Villaver2007, Mustill2012, Rao2018, Ronco2020, Veras2024} and they have been applied to understand the presence of observed white dwarf exoplanets \citep{OConnor2021, Lagos2021}. Advancements in a wide range of detection techniques have allowed seven exoplanets to be observed orbiting white dwarfs \citep{Sigurdsson2003, Luhman11, Gansicke_2019, Vanderburg20, Blackman2021, Mullally2024}, and future surveys and missions have the potential to start classifying the types of planets orbiting white dwarfs and their locations \citep{vanSlujis2018, Morris2021, Limbach2024}. JWST has proven to be a valuable tool for this job as it has been used to detect Jupiter analogs through direct imaging \citep{Mullally2024}, a white dwarf companion from mid-infrared observations using MIRI \citep{Venner2024}, and the MEOW survey is already returning unprecedented results with an exoplanet candidate reported having a separation between 0.1-2 AU \citep{Limbach2024}. 

The small size of white dwarfs could also be a major factor in the difficulty in observing these objects and their accompanying exoplanets as it means that their luminosity is low even with a surface temperature comparable to the Sun. Most white dwarfs have a mass of $0.6 M_\odot$ \citep{Kepler2007, Kleinman2013}, which is found to correspond to a radius of $1.36 R_\oplus$ from \citet{Suh_2000}. Although this creates a large transit depth even for Earth-sized planets, it also means a low transit probability that favors larger planets. Further investigation of these large bodies resembling Jupiter can reveal the possibility of Earth-like planets orbiting close to the star. This question is intriguing as the low luminosity would necessitate that the habitable zone be close to the star \citep{agol2011transit, barnes2013habitable,Loeb13}, but the post-main-sequence evolution may clear out this region. Resolving this problem is a major challenge, making it useful to evaluate the potential for an Earth-like planet in this region to be habitable and predict the environment with the highest chance of successful detection.

The characteristics of the host star play a major role in determining the habitability of a planet, and this is especially true for white dwarfs, as their cooling adds a level of complexity. The habitable zone is classically defined as the region in which liquid water could be sustained on the surface of the planet \citep{Dole1964, Kasting93, Kopparapu_2013, Ramirez2018, Lingam_2021_history}, and it has been studied previously for white dwarfs \citep{agol2011transit, barnes2013habitable, Loeb13,Becker23,Zhan2024}. The habitable zone resides close to the white dwarf and drifts inward as it ages, on account of the cooling associated with the white dwarf; this behavior is depicted in Figure \ref{fig:HZ_comp}, and is discussed subsequently in the paper, where the TRAPPIST-1 system is over plotted for reference.

In order to get a more complete view of the habitability of the planet, a method strongly akin to studying the habitable zone can predict the region at which a planet must orbit to allow for photosynthesis and pre-biotic photochemical reactions. Both processes contribute directly to the formation and sustainability of life making them good indicators of a planet's habitability, and they are dependent on the radiation received from the host star. Photosynthesis is known to have significantly altered the biosphere of Earth, shaping the evolutionary track of all life as well as the geochemical landscape \citep{Knoll_2017}. The photochemical reactions that potentially led to the origin of life utilized ultraviolet (UV) light, specifically UV-C radiation, as a primary source of energy \citep{MA24}. Past studies have focused on the flux received in various wavelength ranges that would fall within these regimes leading to a promising outlook for the potential for life and its detection \citep{Fossati_2012, Loeb13, Kozakis2018, Kaltenegger20}. Also, this framework for evaluating habitability based on these parameters has been used to gain meaningful insight into the outlook for life on habitable planets around brown dwarfs by \citet{lingam2020prospects}. 

This paper looks to evaluate the habitability of an Earth-like planet orbiting a white dwarf based on a novel synthesis of three facets: the habitable zone, photosynthesis, and UV-driven abiogenesis. We begin by defining the habitable zone as a function of the white dwarf's age in Section \ref{sec:HZ}, and verify that it agrees with the findings of previous studies. In Section \ref{sec:hab} the orbital regions at which the planet can sustain photosynthesis (Section \ref{subsec:photo}) and UV-mediated pre-biotic chemistry (Section \ref{subsec:abio}) are calculated. The implications of these results are discussed in Section \ref{sec:discussion} in the context of the potential that life can be found in such a system and gauging whether it is possible to detect with current technology.

\section{Time-varying Habitable Zone for White Dwarfs} \label{sec:HZ}

In the following sections we compute the habitable zone for an Earth-like planet orbiting a white dwarf, and explore the consequences of their time-dependent luminosity. Relevant equations and constants used are listed and defined in Appendix \ref{ap:eqs}.

\subsection{Stellar Temperature} \label{subsec:temp}
It is expected that for life to be able to exist on a planet, the energy received by the planet must be in the proper range to allow for liquid water to permanently remain on the surface, namely, it must dwell in the habitable zone. Since white dwarfs cool as they age, this zone will migrate inward at a rate that corresponds to the cooling rate of the star. The luminosity of white dwarfs with mass of about $0.55 M_\odot \leq M_{\text{WD}} \leq 0.65 M_\odot$ is described in the cooling function from \citet{barnes2013habitable} shown below, where $L$ is the base-10 logarithm luminosity in units of $L_\odot$ and $t$ is time in Gyr:
\begin{equation}\label{eq:cooling}
    L = -2.478-0.7505t+0.1199t^2-6.686\times10^{-3}t^3.
\end{equation}

The effective temperature $(T_{\text{WD}}\text{[K]})$ of the white dwarf can be computed from its luminosity as follows 
\begin{equation} \label{eq:T_wd}
    T_{\text{WD}}=6.03\times10^4 \text{ K } \times \left(\frac{R_{\text{WD}}}{R_{\oplus}}\right)^{-1/2}\left(\frac{L_{\text{WD}}}{L_{\odot}}\right)^{1/4},
\end{equation}
where $R_{\text{WD}}$ is the radius of the white dwarf, $R_\oplus$ is Earth's radius, $L_{\text{WD}}$ is the luminosity of the white dwarf (which is a function dependent on its age), and $L_\odot$ is the luminosity of the Sun.

A typical white dwarf has a mass of $0.6 M_\odot$ and a radius of $1.36 R_\oplus$ \citep{pinsonneault2023stellar}, which are the canonical values that will be used here. With these parameters remaining mostly constant through the lifetime of the white dwarf, the habitable zone will only depend on its luminosity and, therefore, its age. 

\subsection{Planetary \& Orbital Elements} \label{subsec:orbital}
Other important orbital factors that must be considered are the eccentricity of the orbit and the Roche limit. A circular orbit for the planet is assumed, as the circularization time depends highly on the orbital radius, and orbits of $\sim 0.01$ AU will likely have circularization times on the order of 10-100 Myr \citep{Matsumara_2008,Pierrehumbert2019}. However, at larger orbital distances, highly eccentric orbits would likely extend the habitable lifetime of Earth-like planets as explained by \citet{Becker23}, which is an additional effect not considered in our leading-order analysis. 

The Roche limit $(d_{\text{RL}})$ must also be determined, as it marks the distance at which tidal disruption can affect the planet. This ultimate lower bound on the orbital radius is defined as \citep[Section 4.8]{Murray_1999}: $d_{\text{RL}}\approx2.46R_{\star} (\rho_\star /\rho_{\text{p}})^{1/3}$, where $R_{\star}$ is the radius of the star and $\rho_\star$ and $\rho_{\text{p}}$ are the density of the star and the planet, respectively. Upon rearranging, we have
\begin{equation} \label{eq:roche_wd}
    d_{\text{RL}}\text{[AU]}\approx7.267\times10^{-3}\times \left(\frac{\rho_{\text{p}}}{\rho_{\oplus}}\right)^{-1/3}\left(\frac{M_{\text{WD}}}{M_{\odot}}\right)^{1/3},
\end{equation}
which now is dependent on the mass of the white dwarf and the density of the planet. With a white dwarf mass of $0.6 M_\odot$ and an Earth-like planetary density of $5.5 \mathrm{ g/cm^3}$ the Roche limit will remain $6.129\times10^{-3}$ AU throughout our models, and any orbit below this distance will not be habitable. 

The planetary equilibrium temperature is given by $T_{\text{p}}[K]=T_{\text{WD}}\sqrt{R_{\text{WD}}/2a} \times (1-A_{\text{p}})^{1/4}$, where $A_{\text{p}}$ is the Bond albedo, and $a$ is the planet's orbital radius; this expression can be simplified by inputting (\ref{eq:T_wd}) for $T_{\text{WD}}$, and using Earth's albedo of $A_{\text{p}} = 0.3$. We find that the orbital radius $a$ at which the planet would have an equilibrium temperature of $T_{\text{p}}$ is:
\begin{equation} \label{eq:HZ}
    a [\text{AU}]=\frac{64818}{T_{\text{p}}^2}\times \left(\frac{L_{\text{WD}}}{L_{\odot}}\right)^{1/2},
\end{equation}
which is used for calculating the limits of the habitable zone, as sketched in the next section.

\subsection{Habitable Zone Limits} \label{subsec:HZ_limits}
\begin{figure}
    \centering
    \includegraphics[scale=0.55]{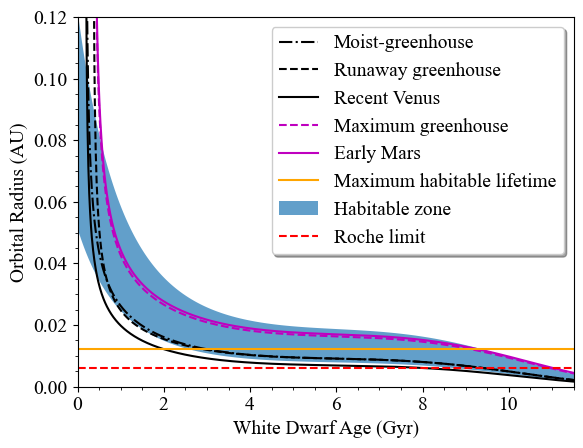}
    \caption{Shaded blue region shows a simplified habitable zone (HZ) explained in Section \ref{subsec:HZ_limits}. This is compared against the standard approach of determining the HZ limits based on the stellar flux (described in Section \ref{subsec:HZ_limits}): inner limits of the HZ are the moist-greenhouse (black dash-dotted), runaway greenhouse (black dashed), and recent Venus (black solid) thresholds, whereas outer limits are the maximum greenhouse (magenta dashed) and early Mars (magenta solid) thresholds. Solid orange line marks the orbital radius at which the interval spent in the temporally changing habitable zone is maximized.}
    \label{fig:WD_FluxHZ}
\end{figure}

The standard approach of computing the habitable zone boundaries involves the incident stellar flux \citep{Kasting93,Ramirez2018}, as described shortly. However, a ``shortcut'' for calculating the habitable zone solely in terms of the equilibrium planetary temperature was proposed in \citet{kaltenegger2011exploring}; the inner and outer boundaries correspond to $T_{\text{p}}$ of 270 K and 175 K, respectively. We proceed with this simpler model hereafter, since it is easy to work with, and it accurately reproduces the results of the standard approach, as illustrated in Figure \ref{fig:WD_FluxHZ} and outlined below.

The ``shortcut'' results are plotted in Figure \ref{fig:WD_FluxHZ}, with the habitable zone being the region shaded in blue and the Roche limit being the dashed red line. This plot reveals how the habitable zone as a whole migrates inward and changes in width as the white dwarf ages. While the inner limit of the habitable zone is close to the Roche limit, it does not have any impact on the habitable zone until an age of nearly 9 Gyr. Also, there is a period when the white dwarf's age is between $\sim2$ Gyr and $\sim9$ Gyr that the habitable range appears to stabilize.

The aforementioned standard approach for calculating the inner and outer edges of the habitable zone is explained in \citet{Kopparapu_2013}. This formalism uses suitable critical stellar fluxes (e.g., for moist greenhouse) to determine the possible inner edges, and the climate conditions of a maximum greenhouse and early Mars to define possible outer edges of the habitable zone. The effective stellar flux and orbital distance at which it occurs is calculated from the following equations, which are taken directly from \citet{Kopparapu_2013}:
\begin{equation}\label{eq:S_eff}
    S_{\text{eff}} = S_{\text{eff}\odot}+aT_{\star}+bT_{\star}^2+cT_{\star}^3+dT_{\star}^4,
\end{equation}
\begin{equation}\label{eq:flux_radius}
    a [\text{AU}] = \left(\frac{L/L_{\odot}}{S_{\text{eff}}}\right)^{1/2}.
\end{equation}
Note that $S_{\text{eff}}$ is now interpreted as the effective flux of the white dwarf (i.e., a stellar remnant and not a main-sequence star), which is a function of its (time-dependent) luminosity and temperature, namely, $T_{\star} = T_{\text{WD}}$, and $L = L_{\text{WD}}$. At this stage, note that equation (\ref{eq:S_eff}) was formulated for $T_{\star} < 7200$ K \citep{Kopparapu_2013,AFK22}. Hence, according to equations (\ref{eq:cooling}) and (\ref{eq:T_wd}), the preceding constraint pertaining to the white dwarf temperature suggests that our subsequent findings are likely to exhibit higher accuracy when the white dwarf has an age of $\gtrsim 1.5$ Gyr.

Substituting the cooling function of equation (\ref{eq:cooling}) in equation (\ref{eq:flux_radius}) yields the inner and outer edges of the habitable zone depending on the chosen coefficients. These coefficients in equation (\ref{eq:S_eff}) are defined by the limits under consideration. The inner limits correspond to the moist greenhouse, runaway greenhouse, and recent Venus thresholds, whereas the outer limits are based on the maximum greenhouse and early Mars thresholds. These inner/outer limits were derived in \citet{Kopparapu_2013} for various circumstances under which the planetary environment could allow for liquid water to exist under different assumptions for either theoretical or empirical bottlenecks on habitability.

As mentioned before, Figure \ref{fig:WD_FluxHZ} demonstrates how the standard approach of computing the habitable zone matches well with the simpler model using equilibrium planet temperature. Moreover, we perform several consistency checks of the simpler model toward the end of Section \ref{subsec:HL} by comparing results with other publications. Hence, this simple model will be used in the following calculations and analysis.

\subsection{Habitable Lifetime} \label{subsec:HL}
Since the rate at which the habitable zone is moving inward is not constant, it is helpful to define the habitable lifetime as the amount of time that any fixed orbital radius will remain in the habitable zone (which changes temporally). This can be interpreted as the interval that passes from the initial moment the planet (at that radius) enters the habitable zone until it finally leaves the habitable zone. Equation (\ref{eq:HZ}) can be used to estimate the white dwarf age as a function of the inner and outer limits of the habitable zone, and subtracting the two functions provides the habitable lifetime for all values of $a$ relevant to this habitable zone. The solid orange line in Figure \ref{fig:WD_FluxHZ} shows the orbital distance at which the planet will remain habitable the longest being $\sim0.012$ AU, which has a habitable lifetime of nearly 7 Gyr.

For comparison, it is estimated that the total habitable lifetime of Earth is likely about 6 Gyr \citep{Caldeira1992, Goldblatt_2012, Rushby_2013, Wolf_2015, deSousa_2020}, and the timescales required for most major evolutionary events are on billion-year timescales \citep{Knoll_2017, Lingam_Loeb_2021}. To maximize the habitable lifetime, the planet would ideally have arrived or formed at a stable orbit near the peak of 0.012 AU before this distance enters the habitable zone, and if the atmosphere contains extensive clouds or develops a strong enough greenhouse effect, the habitable zone could be expanded \citep{Kasting93, Pierrehumbert_2011, Vladio_2013, Windsor_2024}.

M-dwarfs are the stellar type with the longest expected habitable lifetime, as they have the longest main-sequence lifetime of any star. This means that their effective temperature and effective flux could remain constant for up to trillions of years \citep{Shields2016}. Perhaps the most well known and exciting M-dwarf in the fields of exoplanets and astrobiology is TRAPPIST-1, as it has seven planets resembling the Earth in size, three of which are within its habitable zone \citep{Gillon2017}. Figure \ref{fig:HZ_comp} shows how the time-varying habitable zone (blue region) of a white dwarf at three moments in time (white dwarf ages of 2, 6, and 10 Gyr) compares with its counterpart for TRAPPIST-1 (orange region). 

These ages are chosen to illustrate how the habitable zone changes over time, migrating progressively inward at greater ages. The three snapshots correspond to epochs for which the orbital distance of 0.012 AU -- corresponding to the longest habitable lifetime of an Earth-like planet -- would be just inside the inner edge of the habitable zone, within the habitable zone, and just outside the outer edge of the habitable zone, respectively. Although a $0.6 M_\odot$ white dwarf has a higher effective temperature than TRAPPIST-1, its much smaller radius leads to a habitable zone that is closer in and less broad when compared to the cool M-dwarf. In this particular respect the M-dwarf stars have an advantage in their ability to host life, but white dwarfs likely emit higher fluxes of photons in biologically relevant wavelength ranges which will be shown in Section \ref{sec:hab}.

The habitable zone calculated here is consistent with prior studies. \citet{agol2011transit}, using a different white dwarf cooling function and method for calculating $T_{\text{p}}$, predicted a maximum habitable lifetime of 8 Gyr at an orbital distance of $\sim0.01$ AU and a minimum habitable lifetime of 3 Gyr for the range $d_{RL} < a < 0.02$ AU, which both agree with our findings. Likewise, the methodology of \citet{barnes2013habitable} results in a habitable zone of $0.015 \text{ AU}\lesssim a \lesssim 0.03$ AU for a 2 Gyr white dwarf, and our calculation gives a habitable zone of $0.0149 \text{ AU }\leq a \leq 0.0354$ AU for the same age. 

For a white dwarf age of 2 Gyr, the radiation-only model of \citet{Becker23} gives a habitable zone for a $0.54 M_\odot$ white dwarf of $0.025 \text{ AU} \lesssim a \lesssim 0.050$ AU which is comparable to our results considering the slightly lower mass. Also, when including tidal heating \citet{Becker23} finds the habitable lifetime to be between $6-10$ Gyr in agreement with both our result of nearly 7 Gyr, and the expectation of tidal heating to increase the habitable lifetime.

\section{Habitability of Earth-like planet around a White Dwarf}\label{sec:hab}
The decreasing effective temperature of a white dwarf also means that the radiation it emits will shift to longer wavelengths. This ongoing change in the wavelength of radiation received by the planet becomes important when considering the (pre)biotic processes occurring on its surface that allow for the formation and existence of life. The two processes examined below are photosynthesis and UV-mediated prebiotic chemistry, as they are potentially two of the most influential stellar radiation-driven processes contributing to the existence of life \citep{Lingam_Loeb_2021}. It is useful to build upon the habitable zone by including the region at which both liquid water can exist and sufficient energy is received on the surface of the planet to support life. 

The radiation flux incident on a planet $(\Phi[\text{photons }m^{-2}s^{-1}])$ 
\begin{equation} \label{eq:phi}
    \Phi=\frac{\dot{N}_{\text{WD}}}{4\pi a^2},
\end{equation}

\noindent depends on the orbital radius $a$ [AU] and on the photon flux emitted by the white dwarf ($\dot{N}_{\text{WD}}$ [photons $s^{-1}]$) over the wavelength range of $\lambda_{\text{min}} \leq \lambda \leq \lambda_{\text{max}}$ [m].
\begin{equation} \label{eq:N}
    \dot{N}_{\text{WD}}=4\pi R_{\text{WD}}^2 \int_{\lambda_{\text{min}}}^{\lambda_{\text{max}}} \frac{2c}{\lambda^4} \left[ \exp\left(\frac{hc}{\lambda k_B T_{\text{WD}}}\right)-1 \right]^{-1} \,d\lambda
\end{equation}

Substituting (\ref{eq:T_wd}) and $1.36 R_\oplus$ in for $T_{\text{WD}}$ and $R_{\text{WD}}$, and combining (\ref{eq:N}) with (\ref{eq:phi}) can provide an expression for the maximal orbital radius for a specific wavelength range as shown in the following sections for both photosynthesis and UV radiation. 

It should be noted that we focus on estimating the maximal orbital radius up to which a given terrestrial exoplanet can sustain a particular process. Hence, as our emphasis is on roughly calculating the upper bound, we do not take into account the effects of atmospheric absorption and scattering, which would serve to reduce the actual threshold for the orbital radius.

\subsection{Photosynthesis}\label{subsec:photo}

The wavelengths used by oxygenic photosynthesis are based on the mechanism used by the photosystem to absorb light, which could possibly adapt to the environment through evolution. Here we will use the conventional range of photosynthetically active radiation (PAR) on Earth, which has a maximum wavelength of 750 nm and a minimum wavelength of 400 nm \citep{Kiang_2007, Lingam_Loeb_2021}. This framework can be easily modified to account for other wavelength ranges such as those considered in \citet{Kiang_2007, Lehmar2021, Lingam_2021_photo}. Using this range also assumes that the atmosphere is optically thin over this range. Therefore the incident PAR flux on the planet is found from equations \ref{eq:phi} and \ref{eq:N} using the wavelength range specified above. The expression to find this flux for a white dwarf as a function of $a$ is: 
\begin{equation}\label{eq:PAR_flux}
\begin{aligned}
    \Phi = 8.0116 &\times10^{19} \times  \left( \frac{a}{1\text{AU}} \right)^{-2} \left( \frac{R_{\text{WD}}}{R_{\oplus}} \right)^{-1/2}\\ &\times \left( \frac{L_{\text{WD}}}{L_{\odot}} \right)^{3/4} \mathcal{F}(t_{\text{WD}},R_{\text{WD}}) ,
\end{aligned}
\end{equation}

\noindent where $\mathcal{F}(t_{\text{WD}},R_{\text{WD}})$ is the function
\begin{equation}\label{eq:F}
    \mathcal{F}=\int_{Y_{\text{min}}}^{Y_{\text{max}}} \frac{y'^2 \,dy'}{e^{y'}-1} ,
\end{equation}

\noindent with $t_{\text{WD}}$ being the age of the white dwarf in Gyr and the limits of integration ($Y_{\text{min}}$ and $Y_{\text{max}}$) defined as
\begin{equation}\label{eq:Ymin}
    Y_{\text{min}}=0.3181 \times \left( \frac{R_{\text{WD}}}{R_{\oplus}} \right)^{1/2} \left( \frac{L_{\text{WD}}}{L_{\odot}} \right)^{-1/4} ,
\end{equation}
\begin{equation}\label{eq:Ymax}
    Y_{\text{max}}=0.5964 \times \left( \frac{R_{\text{WD}}}{R_{\oplus}} \right)^{1/2} \left( \frac{L_{\text{WD}}}{L_{\odot}} \right)^{-1/4}.
\end{equation}
A detailed procedure explaining the construction of this formula can be found in \citet{lingam2020prospects} where it is applied to brown dwarfs. This method has also been used by \citet{Hall2023} when defining a photosynthetic habitable zone.

The PAR flux will decrease for greater orbital distances and as the white dwarf ages. The minimum flux required by photosynthetic life, or critical PAR flux, is $\Phi_c \approx 1.2\times10^{16} \mathrm{m^{-2} s^{-1}}$ \citep{WOLSTENCROFT2002,MA24}. Substituting $\Phi_c$ for $\Phi_{\text{PAR}}$ and rearranging (\ref{eq:PAR_flux}) to solve for $a$ shows the maximum orbital distance at which a planet would receive enough light in the proper wavelength range for photosynthesis. 
\begin{equation}\label{eq:a_PAR}
    a=81.71 \times \left( \frac{R_{\text{WD}}}{R_{\oplus}} \right)^{1/4} \left( \frac{L_{\text{WD}}}{L_{\odot}} \right)^{3/8} \sqrt{\mathcal{F}(t_{\text{WD}},R_{\text{WD}})}
\end{equation}

\begin{figure}
    \centering
    \includegraphics[scale=0.55]{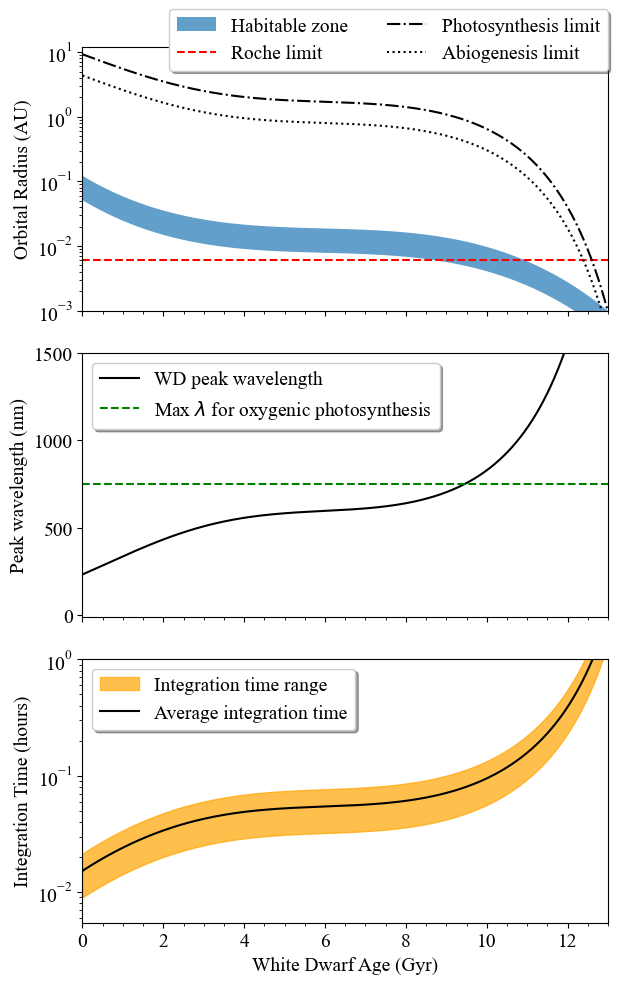}
    \caption{(Top Panel) The orbital radius ($a$ [AU]) at which the flux of photosynthetically active photons and UV radiation reaches their critical flux of $1.2\times10^{16} \mathrm{m^{-2} s^{-1}}$ and $5.44\times10^{16} \mathrm{m^{-2} s^{-1}}$ respectively \citep[Chapter~4]{Lingam_Loeb_2021}. Boundaries for the critical PAR and UV fluxes are the dash-dotted and dotted lines, respectively. Orbital distances below these curves may receive the appropriate flux to support photosynthetic life or abiogenesis. (Middle Panel) The peak wavelength ($\lambda_{\text{peak}}$ [nm]) of a white dwarf's spectral radiation flux as a function of its age (Gyr); cooling of the white dwarf leads to a shift toward longer wavelengths. The green dashed line is the maximum wavelength used for photosynthesis on Earth (750 nm). (Bottom Panel) JWST in-transit integration time needed to achieve a S/N of 5 using transmission spectroscopy to observe biosignatures present in an Earth-like atmosphere, calculated using equation (\ref{SNTranSpec}).}
    \label{fig:WD_PAR_UV_peak_SNR}
\end{figure}

The top panel of Figure \ref{fig:WD_PAR_UV_peak_SNR} shows the resulting curve plotted along with the habitable zone with any point below the dash-dotted line receiving enough PAR flux. It is promising that the habitable zone is entirely within the critical PAR zone. This implies that photosynthetic life would not be limited by any stellar factors. 

As stated earlier, it is possible that the mechanisms used for photosynthesis could adapt to their specific environment, and the similarity between the peak of the Sun's spectral flux and the peak absorbance of photosynthesis on Earth supports this idea \citet{Lingam_2021_photo}. The peak spectral flux can be easily calculated from Wien's displacement law resulting in the relationship between a white dwarf's age and its peak wavelength ($\lambda_{\text{peak}}$ [nm]) given by
\begin{equation}\label{eq:peak}
    \lambda_{\text{peak}}=48.10 \times \left( \frac{R_{\text{WD}}}{R_{\oplus}} \right)^{1/2} \left( \frac{L_{\text{WD}}}{L_{\odot}} \right)^{-1/4}.
\end{equation}

This reveals that the peak wavelength during the white dwarf's habitable lifetime is nearly identical to that of the Sun ($635$ nm), and below the maximum wavelength that can be used by photosynthesis on Earth as seen in the middle panel of Figure \ref{fig:WD_PAR_UV_peak_SNR}. It also lends credence to the notion that it would be possible for the same mechanisms of photosynthesis to evolve in a white dwarf planetary system supporting the choice of wavelength parameters in the PAR flux calculation.

\subsection{Abiogenesis}\label{subsec:abio}

Another process that can contribute to the habitability of a planet and is dependent on the star it orbits is prebiotic chemical reactions that require ultraviolet (UV) radiation. While there is not a definite answer yet for the likelihood and mechanisms of life's origination \citep{Lingam2024}, UV radiation seems to be one of the primary energy sources able to drive the chemical reactions of abiogenesis. Experiments have determined that the required wavelength range to make such reactions efficient is $200 \text{nm} < \lambda < 280 \text{nm}$, with a minimum photon flux of $5.44\times10^{16} \mathrm{m^{-2} s^{-1}}$ \citep{Rimmer_2018, Lingam_Loeb_2021}. This critical UV flux is defined by the limit at which the photochemistry reactions begin to have quicker reaction rates than the "dark" reactions that proceed without the presence of UV radiation \citep{Rimmer_2018}. This is required for the UV-mediated abiogenesis reactions comprising the crux of the so-called "cyanosulfidic protometabolism" to be favored over other reactions.

Determining the outer limit at which a planet would receive the minimum flux required to facilitate UV-mediated prebiotic chemistry follows the same method used for PAR but with the different wavelength range mentioned above. The UV flux is therefore found from (\ref{eq:phi}) and (\ref{eq:N}) with $\lambda_{\text{max}} = 280$ nm and $\lambda_{\text{min}} = 200$ nm, and used to find the orbital distance that receives the critical flux:
\begin{equation}\label{eq:a_UV}
    a=38.375 \times \left( \frac{R_{\text{WD}}}{R_{\oplus}} \right)^{1/4} \left( \frac{L_{\text{WD}}}{L_{\odot}} \right)^{3/8} \sqrt{\mathcal{G}(t_{\text{WD}},R_{\text{WD}})}.
\end{equation}

with the integration function $\mathcal{G}(t_{\text{WD}},R_{\text{WD}})$ as follows:

\begin{equation}\label{eq:G}
    \mathcal{G}=\int_{Z_{\text{min}}}^{Z_{\text{max}}} \frac{z'^2 \,dz'}{e^{z'}-1}
\end{equation}

where the limits of integration ($Z_{\text{min}}, Z_{\text{max}}$) are

\begin{equation}\label{eq:Zmin}
    Z_{\text{min}}=0.8519 \times \left( \frac{R_{\text{WD}}}{R_{\oplus}} \right)^{1/2} \left( \frac{L_{\text{WD}}}{L_{\odot}} \right)^{-1/4},
\end{equation}
\begin{equation}\label{eq:Zmax}
    Z_{\text{max}}=1.1927 \times \left( \frac{R_{\text{WD}}}{R_{\oplus}} \right)^{1/2} \left( \frac{L_{\text{WD}}}{L_{\odot}} \right)^{-1/4}.
\end{equation}

Equation (\ref{eq:a_UV}) as a function of the white dwarf's age, shown in the top panel of Figure \ref{fig:WD_PAR_UV_peak_SNR} as the dotted black line, reveals the outer edge of the abiogenesis zone. Any orbital distance below this line will receive the required flux to support the necessary photochemical reactions for abiogenesis. It is apparent that the entire habitable zone falls within the zone that allows for efficient UV-mediated abiogenesis. For this range of UV radiation there is no maximum flux limit, but it is possible that too high of a UV flux in other wavelength regions could negatively impact habitability through processes such as atmospheric escape \citep{Lammer2013}.

\section{Discussion}\label{sec:discussion}

Searching for life beyond Earth has advanced substantially with the progress of theoretical modeling and observational capabilities. Continuing to advance the field will partially rely on identifying the best environment for life to form, but without narrowing the search too much. In the past this has meant that Sun-like stars were prioritized as they are the most familiar target. Here we discuss how white dwarfs offer a unique opportunity as observational targets and hosts of life.

\subsection{Outlook for Observation}
White dwarf exoplanet detection is growing as an exciting field with the potential to advance our understanding of planet formation, exoplanet atmospheres, planetary system dynamics, and the fate of the solar system. The variety of exoplanets orbiting white dwarfs have been detected using a different method showing that these systems are pushing the boundaries of current capabilities. The size of a typical white dwarf gives transit detection an advantage, which is beneficial for biosignature detection as it is a powerful means to determine the composition of an exoplanet's atmosphere. For JWST parameters, including a telescope aperture of $D = 6.5$ m and a total throughput of $\xi = 0.4$ \citep{Cowan2015}, the simplified limit of the signal-to-noise ratio (S/N) for transmission spectroscopy is given by equation (4) of \citet{Fujii_2018}:
\begin{eqnarray}\label{SNTranSpec}
&& \mathrm{S/N} \approx 10\, \left(\frac{N_H}{4}\right)\left(\frac{H_a}{8\,\mathrm{km}}\right)\left(\frac{R}{R_\oplus}\right)\left(\frac{R_\star}{0.1\,R_\odot}\right)^{-1} \\
&& \hspace{0.5in} \times \left(\frac{{n}_\lambda(\lambda;\,T_\star)}{{n}_\lambda(3\,\mathrm{\mu m};\,2500\,\mathrm{K})}\right)^{1/2}\left(\frac{\Delta \lambda}{0.1\,\mathrm{\mu m}}\right)^{1/2} \nonumber \\
&& \hspace{0.5in} \times \left(\frac{\Delta t}{30\,\mathrm{hr}}\right)^{1/2} \left(\frac{d_\star}{10\,\mathrm{pc}}\right)^{-1} \left(\frac{D}{6.5\,\mathrm{m}}\right) \left(\frac{\xi}{0.4}\right)^{1/2}, \nonumber
\end{eqnarray}
where $N_H$ is a dimensionless number that quantifies the strength of spectral features; $H_a$ is the atmospheric scale height; $R_\star$ is the radius of the stellar object (white dwarf in our case), $n_\lambda$ is the photon spectral flux density ($T_\star$ is the white dwarf temperature); $\Delta \lambda$ is the wavelength resolution; $d_\star$ is the distance of the planet from Earth; and $\Delta t$ is the in-transit integration time. This estimate should be considered an idealization because it only incorporates the effect of photon shot noise, and similarly does not account for all interactions of transmitted radiation with the atmosphere.

In the bottom panel of Figure \ref{fig:WD_PAR_UV_peak_SNR}, the average in-transit integration time needed to achieve a S/N of $5$ based on the age of the white dwarf is shown; most parameters are held fixed at their typical values in equation (\ref{SNTranSpec}). It is not until the age of $\sim10$ Gyr that the integration time exceeds $1$ hour. As explained, say, in \citet{Lingam_Loeb_2021}, it can be demonstrated from this formula that lower-mass stars are preferred for generating stronger signals (if all other factors are fixed), and S/N is likewise maximized when observing in wavelengths where the stellar flux peaks. This feature is beneficial for white dwarfs where the lower mass permits stronger signals than a Sun-like star, but a similar effective temperature allows for comparison in the molecular features detected.

This suggests that any planet discovered orbiting a white dwarf should receive substantial investigation as high quality data can be easily obtained. Previous studies have also shown that white dwarfs are excellent candidates for such observations as they can lead to the identification of strong biosignatures; namely N\textsubscript{2}, O\textsubscript{2}, O\textsubscript{3}, CO\textsubscript{2}, and CH\textsubscript{4} \citep{Loeb13, Kozakis20, Kaltenegger20}. These hopes have likely been confirmed as recently there is evidence for the first observation of a white dwarf exoplanet atmosphere from a JWST transmission spectrum of the gas giant WD 1856b \citep{MacDonald2024JWST}. 

\subsection{Formation and Evolution of Life}
Although the potential to detect any present biosignatures is strong, it relies on the planet being habitable in the first place, and our understanding of the various ways life could present itself. Defining the habitable zone of white dwarfs is a good first step in understanding the habitability of white dwarf planetary systems.

Figure \ref{fig:WD_FluxHZ} shows that a major challenge to planetary habitability for such a system is the inward migration of the habitable zone. After the white dwarf is formed, it rapidly cools until it is between 2 and 4 Gyr old, where the cooling slows down and remaining stable until it attains an age of at least 8 Gyr. This creates a range of orbital distances where there are billions of years between the time the planet would enter the habitable zone and when it would leave it, which we refer to as the habitable lifetime. Another promising result from this habitable zone is that its width seems to remain constant as the star ages until the end of its habitable lifetime, which is understood to manifest when it is cut-off by the Roche limit. This period of stability creates a window of time for life to begin and evolve that is very similar in duration to that of Earth's own habitability interval. 

Even considering the various planetary environments of a moist-greenhouse, runaway greenhouse, recent Venus, maximum greenhouse, and early Mars does not have a large impact on the habitable zone across the range of white dwarf ages which see the most stable cooling period as seen in Figure \ref{fig:WD_FluxHZ}. A planet with a thick atmosphere similar to that of Venus could, in principle, slightly extend the habitable lifetime since it would enter the habitable zone earlier.

The top and middle panels of Figures \ref{fig:WD_PAR_UV_peak_SNR} reveal that the necessary conditions for a white dwarf to host life are present in a way that is similar, and in some cases perhaps even superior, to the conditions seen on Earth. The top panel of this figure shows that the expected UV-C flux provides enough energy to allow for abiogenesis, and it is possible that photosynthesis will evolve similar to that seem on Earth because of the physical constraint that the peak wavelength of the white dwarf resembles the Sun. Also, photosynthetic organisms will have the ability to exist for billions of years as a result of the habitable lifetime of planet's orbiting white dwarfs. 

This overlap of the UV abiogenesis zone and the PAR zone in the top panel of Figure \ref{fig:WD_PAR_UV_peak_SNR} is an advantage that white dwarfs have over other low-mass stars. From this perspective white dwarfs offer the "just right" option having longer habitable lifetimes than some G-type stars and possibly with less hurdles to form life than low-mass stars and brown dwarfs. However, it is still important to recognize that the best pathway for abiogenesis is an unknown along with the full extent to which UV-C flux actually contributes toward overcoming such a big obstacle. Atmospheric escape is another obstacle pertaining to high UV flux during the cooling of white dwarfs which still needs to be examined further for rocky planets, but its ability to strip away large envelopes of hydrogen and helium could be beneficial for the presence of liquid water and therefore life \citep{Gallo2024}. 

If life is indeed able to form in these conditions, the long habitable lifetime of white dwarfs would provide the appropriate timescales over which it is expected that technological life can eventually evolve on temperate planets, thereby opening up the possibility of seeking out technosignatures  \citep{Lingam_Loeb_2021} on these worlds in the future. The maximum habitable lifetime indicates that it is possible for a planet orbiting a white dwarf to be habitable for nearly 7 Gyr. Also, there exists mechanisms that can increase the habitable lifetime of a planet. With a slightly eccentric orbit, the tidal heating and circularization could provide an additional energy source and drive the orbit of the planet inward along with the habitable zone for at least a short period of time. These mechanisms were proven to slightly increase the habitable lifetime by \citet{Becker23}. Tidal forces also have the potential to impact the photosynthesis zone if they result in the planet being tidally locked with the white dwarf. This could potentially be a limiting factor for where life is able to survive on the surface of the planet as explained by \citet{Hall2023}.

The orbital distance at which this maximum habitable lifetime exists also provides a tool for understanding which future exoplanet detections should be prioritized when searching for biosignatures. If it is assumed that the formation of life takes at least 1 Gyr, then any Earth-like planet with a constant orbital distance of between the Roche limit ($\sim 0.006$ AU) and $\sim 0.06$ AU could host life. There is evidence for life to have existed on Earth about 0.8 Gyr after it is expected to have been habitable, and technologically intelligent life was present after 3.7 Gyr measured from the earliest evidence for life \citep{Lingam_Loeb_2021}. This timeline fits well within the habitable lifetime of a white dwarf exoplanet, but of course it is highly dependent on the specific environment of the planet and other nearby bodies.

Knowing which stars or regions of the universe are most likely to be inhabited by advanced technological species can help narrow SETI projects for which high sensitivity is integral and only possible with targeted searches \citep{Lingam_Loeb_2021}. Habitable zones that have the most long-term stability are anticipated to offer the best chance for intelligent life to evolve, so understanding which systems create such an environment will improve these targeted searches. In this regard, we comment that M-dwarfs might provide the most stable habitable zones as they can remain on the main-sequence for up to trillions of years and even some of the cooler ones, such as TRAPPIST-1, have habitable zones larger than white dwarfs (Figure \ref{fig:HZ_comp}).

The major advantage that white dwarfs have over M-dwarfs is that their effective temperature creates an environment in which the PAR zone and UV abiogenesis zones overlap, and the peak wavelength emitted is similar to that of the Sun (top and middle panels of Figure \ref{fig:WD_PAR_UV_peak_SNR}). This similarity to what we see on Earth is incredibly useful in searching for extraterrestrial life as it seems reasonable to assume that life-as-we-know-it is conservatively presumed to form preferentially in conditions resembling Earth. Until the formation and evolution of life is better understood it is beneficial to eliminate as many unknown factors as possible. Combining conditions suitable for recognizable evolution with relatively long-term stability makes white dwarfs a promising target in the search for life (e.g., technological or microbial) along with other stellar types like M- and K-dwarfs. 

Although determining the age of white dwarfs is difficult, their luminosity functions can be used to estimate ages of stellar populations. \citet{Kilic_2017} found the ages of the thin disk, thick disk, and halo to be approximately 7 Gyr, 8.7 Gyr, and 12.5 Gyr respectively using luminosity functions. It is expected that after 8 Gyr the percentage of stars that would already have evolved into white dwarfs in the thin disk, thick disk, and halo are $44.3\%$, $73\%$, and $79.5\%$ respectively \citep{Hurley2000}. This implies that nearly all $0.6 M_{\odot}$ white dwarfs are less than 10 Gyr old (i.e., since they became white dwarfs) and within their habitable lifetime. The proportion of white dwarfs will only continue to rise leading to more opportunities for evolved life to exist and be detected. 

The main limiting factor therefore becomes the probability of an Earth-like planet having a stable orbit around a white dwarf at approximately 0.012 AU. While a planet may survive engulfment by its host during the red giant branch (RGB) phase \citep{Merlov2021}, problems could arise with the destabilization of a giant planet in the system and volatile depletion. Post-main-sequence planet formation then becomes a key problem to solve so that habitability of white dwarf systems can be fully understood. Explaining the existence of a planet in the habitable zone of a white dwarf using post-main-sequence formation seems promising as it is expected that all polluted white dwarfs have circumstellar material near their Roche limit \citep{Akshay2024}, and it has been shown that such "exoplanet recycling" is possible by \citet{vanLieshout2018}.

\section{Conclusions}\label{sec:conclusions}

Ultimately, white dwarfs offer an exciting opportunity in the search for life that should become more of a focus as observational techniques continue to advance. Our work offers a number of striking conclusions, some of which do not appear to have been ever quantified before. The effective temperature of a typical white dwarf during its habitable lifetime provides conditions for abiogenesis and photosynthesis similar to those seen on Earth. Moreover, the habitable lifetime could be slightly longer than the Sun's at the ideal orbital distance of 0.012 AU. With a long habitable lifetime and overlap in PAR and abiogenesis zones white dwarfs are an ideal host for the origin and advanced evolution of life.

Of course, the calculations explained above all assume the planet has an Earth-like atmosphere, and deviations from this would change both the inner and outer limits of the habitable zone. Further understanding of how a white dwarf specifically would affect the atmospheric dynamics and evolution of a terrestrial atmosphere will be valuable in understanding habitability, and even the outlook for detection. Higher rotation rates may slightly decrease planetary albedo and lead to more warming \citep{Yang2014}, which could have some benefits with a cooling primary, such as a white or brown dwarf. The slight dependence on atmospheric and environmental conditions is also indirectly captured in Figure \ref{fig:WD_FluxHZ} as there is some variation with different greenhouse conditions. 

While these results show that an Earth-like exoplanet can be habitable around a white dwarf, they may also imply that it will be unlikely to find such a planet. This orbital region would have likely been engulfed during its host's RGB phase, and the orbit of any existent planet may be rendered dynamically unstable. In surveys using \emph{Gaia} data, 130 white dwarfs have been identified \citep{Hollands2018}, and only about 8 planets which will be orbiting between 1.6 and 3.91 AU are anticipated to be detected \citep{Sanderson2022}. However, there seems to be a disconnect between the expected abundance of white dwarf exoplanets and the discoveries made up to this point. It is conceivable that the scarcity of observations is a byproduct of the difficulty that observing white dwarfs presents. It will not be long until this forbidden zone around white dwarfs is better understood, as \citet{Limbach2024} has shown JWST's ability to probe white dwarf planets at intermediate separations.

Although observations will be able to return valuable empirical data, the small radial size of white dwarfs would render transiting exoplanets uncommon during such observations and direct imaging challenging. Rocky exoplanets similar in size to Earth are difficult to detect in ideal conditions, and the size and luminosity of white dwarfs adds to the challenge. A transit probability of approximately 0.006 \citep{Loeb13} means that even if a habitable planet is orbiting at the ideal distance it is statistically unlikely to be transiting. Until methods such as direct imaging or thermal emission spectroscopy are further developed, theoretical analyses of such a planetary system are important.

Exoplanets orbiting white dwarfs seem to provide a uniquely promising environment that may be stable enough over gigayear timescales to permit the origin and evolution of life. Currently, it seems that it would be uncommon for an Earth-like planet to be found orbiting a white dwarf at such close distances. Planet formation from recycled material could alter this view on the basis of the prevalence of white dwarf pollution and circumstellar material. The upside for white dwarf exoplanet habitability and understanding planetary system evolution warrants the attention of future observational missions, which have the potential to empirically test some of the predictions made in this paper.

\section*{Acknowledgements}
    
We would like to thank Dr Howard Chen at the Florida Institute of Technology for his valuable insight and advice. We also thank the anonymous reviewer for providing insightful comments which help to improve the quality of the research developed and clarify the outcomes in the paper. This research utilized NASA's Astrophysics Data System and the Python packages \textsc{matplotlib} \citep{Hunter:2007}, \textsc{numpy} \citep{harris2020array}, and \textsc{scipy} \cite{2020SciPy-NMeth}. C.W. acknowledges the support from the Ortega Observatory to complete this research. P.P. acknowledges funding from the UK Research and Innovation (UKRI) under the UK government's Horizon Europe funding guarantee from ERC (under grant agreement No. 101076489).

\appendix
\section{Equations and Constants} \label{ap:eqs}
The various governing equations and constants used to calculate the necessary expressions in the paper are provided: stellar effective temperature (Section \ref{subsec:temp}); effective flux (Section \ref{subsec:HZ_limits}); Roche limit (Section \ref{subsec:orbital}); habitable zone orbital distances (Sections \ref{subsec:orbital} and \ref{subsec:HZ_limits}); planet equilibrium temperature (Section \ref{subsec:orbital}); critical flux orbital distances (Sections \ref{subsec:photo} and \ref{subsec:abio}); and radiation flux (Section \ref{subsec:photo}). Luminosity is adopted from \citet{barnes2013habitable}.

\begin{sidewaystable}
  \begin{center}
    \begin{tabular}{ccc}
    Variables& Equations& References\\
     \midrule\midrule
Stellar Effective\\Temperature($T_{WD}$)
&   $\begin{aligned} T_{\text{WD}} \text{[K]}=6.03\times10^4\times \left(\frac{R_{\text{WD}}}{R_{\oplus}}\right)^{-1/2}\left(\frac{L_{\text{WD}}}{L_{\odot}}\right)^{1/4}
\end{aligned}$        &   \\
    \cmidrule(l  r ){1-3}
     Luminosity ($L_\text{WD}$) 
     & $\begin{aligned} \log_{10}{\left(\frac{L_\text{WD}}{L_\odot}\right)} = -2.478 - 0.7505\left(\frac{t}{1 \mathrm{Gyr}}\right) + 0.1199\left( \frac{t}{1 \mathrm{Gyr}} \right)^2 - 6.686\times10^{-3}\left( \frac{t}{1 \mathrm{Gyr}} \right)^3 
     \end{aligned}$ & \citep{barnes2013habitable} \\ 
    \cmidrule(l r ){1-3}
     Effective Flux ($S_{\text{eff}}$) 
     & $\begin{aligned}
     &S_\text{eff} [S_\odot] = S_{\text{eff}\odot} + aT_\text{WD} + bT_\text{WD}^2 + cT_\text{WD}^3 + dT_\text{WD}^4
     \end{aligned}$ & \citep{Kopparapu_2013} \\ 
    \cmidrule(l r ){1-3}
Roche Limit ($d_\text{RL}$)
    &   $\begin{aligned}
        &d_\text{RL} \text{[AU]} \approx 7.267\times10^{-3}\times\left( \frac{\rho_\text{p}}{\rho_\oplus} \right)^{-1/3}\left( \frac{M_\text{WD}}{M_\odot} \right)^{1/3} 
        \end{aligned}$       &           \\
    \cmidrule(l r ){1-3}
Orbital Distance($a$)
    &   $\begin{aligned}
        &\text{From $T_\text{p}$ : } a\text{[AU]} = \frac{64818}{T_\text{p}^2}\times\left( \frac{L_\text{WD}}{L_\odot}\right)^{1/2}  \\
        &\text{From $S_\text{eff}$ : } a\text{[AU]} = \left( \frac{L_\text{WD}/L_\odot}{S_\text{eff}} \right)^{1/2} \\
        &\text{PAR : } a\text{[AU]} = 81.71\times \left( \frac{R_\text{WD}}{R_\odot} \right)^{1/4}\left( \frac{L_\text{WD}}{L_\odot} \right)^{3/8} \sqrt{\mathcal{F}(t_\text{WD}, R_\text{WD})} \\
        &\text{UV : } a\text{[AU]}=38.375 \times \left( \frac{R_{\text{WD}}}{R_{\oplus}} \right)^{1/4}\left( \frac{L_{\text{WD}}}{L_{\odot}} \right)^{3/8} \sqrt{\mathcal{G}(t_{\text{WD}},R_{\text{WD}})}
        \end{aligned}
        $       & \citep{Kopparapu_2013, lingam2020prospects} \\
    \cmidrule(l r ){1-3}
Radiation Flux($\Phi$)
    &   $\begin{aligned}
        \text{PAR: } & \Phi_\text{PAR} \text{[photons m$^{-2}$ s$^{-1}$]} = 8.0116\times10^{19}\left( \frac{a}{1 AU} \right)^{-2} \left( \frac{R_\text{WD}}{R_\odot} \right)^{-1/2} \left( \frac{L_\text{WD}}{L_\odot} \right)^{3/4} \sqrt{\mathcal{F}(t_{\text{WD}},R_{\text{WD}})}\\
        \text{UV: } & \Phi_\text{UV} \text{[photons m$^{-2}$ s$^{-1}$]} = 8.0116\times10^{19}\left( \frac{a}{1 AU} \right)^{-2} \left( \frac{R_\text{WD}}{R_\odot} \right)^{-1/2} \left( \frac{L_\text{WD}}{L_\odot} \right)^{3/4} \sqrt{\mathcal{G}(t_{\text{WD}},R_{\text{WD}})}
        \end{aligned}
        $       & \citep{lingam2020prospects} \\
    \cmidrule(l r ){1-3}
Planet Equilibrium\\Temperature ($T_\text{p}$)
    &   $\begin{aligned}
        &T_\text{p} \text{[K]} = T_\text{WD} \sqrt{\frac{R_\text{WD}}{2a}} \left( 1 - A_\text{p} \right)^{1/4}
        \end{aligned}
        $      &     \\
    \midrule\midrule
Constants
    &   $\begin{aligned}
        &\text{Stellar Mass: } M_\text{WD} = 0.6 M_\odot \qquad 
        \text{Stellar Radius: } R_\text{WD} = 1.36 R_\oplus \qquad 
        \text{Planet Bond Albedo: } A_\text{p} = 0.3 \\
        &\text{Planet Radius: } R_\text{p} = R_\oplus \qquad 
        \text{Planet Mass: } M_\text{p} = M_\oplus \qquad
        \text{Planet Density: } \rho_\text{p} = \rho_\oplus 
        \end{aligned}
        $       &   \\
        \midrule
    \end{tabular}
\end{center}
\end{sidewaystable}

\newpage
\bibliography{ref}{}
\bibliographystyle{aasjournal}

\end{document}